\documentclass[letterpaper]{article} 
\usepackage{aaai24}  
\usepackage{times}  
\usepackage{helvet}  
\usepackage{courier}  
\usepackage[hyphens]{url}  
\usepackage{graphicx} 
\usepackage{natbib}  
\usepackage{caption} 
\usepackage{algorithm}
\usepackage{algorithmic}
\usepackage{multirow}
\nocopyright

\usepackage{newfloat}
\usepackage{listings}
\DeclareCaptionStyle{ruled}{labelfont=normalfont,labelsep=colon,strut=off} 
\floatstyle{ruled}
\newfloat{listing}{tb}{lst}{}
\floatname{listing}{Listing}
\title{ArguMentor: Augmenting User Experiences with Counter-Perspectives}
\author{
    Priya Pitre\textsuperscript{\rm 1},
    Kurt Luther\textsuperscript{\rm 1},
}
\affiliations{
    \textsuperscript{\rm 1}Virginia Tech\\
    priyapitre@vt.edu, 
    kluther@vt.edu 
}

\usepackage{bibentry}

\begin{document}

\maketitle

\begin{abstract}
We encounter arguments everyday in the form of social media posts, presidential debates, news articles, and even advertisements. A ubiquitous, influential example is the opinion piece (op-ed). Opinion pieces can provide valuable perspectives, but they often represent only one side of a story, which can make readers susceptible to confirmation bias and echo chambers. Exposure to different perspectives can help readers overcome these obstacles and form more robust, nuanced views on important societal issues. 
We designed ArguMentor, a human-AI collaboration system that highlights claims in opinion pieces, identifies counter-arguments for them using a LLM, and generates a context-based summary of based on current events. It further enhances user understanding through additional features like a Q\&A bot (that answers user questions pertaining to the text), DebateMe (an agent that users can argue any side of the piece with) and highlighting (where users can highlight a word or passage to get its definition or context). Our evaluation on news op-eds shows that participants can generate more arguments and counter-arguments and display higher critical thinking skills after engaging with the system. Further discussion highlights a more general need for this kind of a system.  
\end{abstract}

\section{Introduction}
People encounter and process arguments in a variety of forms every day. Argumentation is a core part of daily decision-making, as we constantly weigh different perspectives and challenge assumptions to reach well-informed conclusions. Engaging in debate and resolving it allows us to refine our thoughts, test the strength of our reasoning, and understand diverse viewpoints. This process is essential for making better decisions, as it encourages critical thinking and helps us navigate complex issues. 

Opinion pieces in the news are one form of this type of debate. An \textit{opinion piece} (or op-ed) is an article where the writer expresses their opinion on political, social, and societal issues. 
These op-eds often have one characteristic in common: they only represent one side of any story. Coppock et al.~\cite{QJPS-16112} find that op-eds are persuasive to both the general public and elites, influencing opinions across society on major issues.
Druckman~\cite{doi:10.1080/10584600500311394} shows that op-eds can inform major election decisions amongst voters as well. 
Considering the widespread popularity and significant impact of opinion pieces in our lives, it becomes crucial to address the potential challenges that may emerge while engaging with such content.

First, humans are susceptible to confirmation bias \cite{10.7554/eLife.71768, Suzuki2021CharacterizingTI}. Most people struggle with confirmation bias, whether they consciously realize it or not \cite{doi:10.1037/1089-2680.2.2.175}.
When reading opinion pieces, people often only read content that aligns with their preconceived notions. People naturally gravitate towards content that aligns with their interests and beliefs. However, the concern arises when such one-sided content is all that people consume, they may lack more complete information that would lead them to reach a different conclusion. 

Second, online information, such as news articles and social media posts, is often designed to exploit biases and create divides, driving more clicks and furthering specific agendas \cite{InformationWars}. As a result, it becomes nearly impossible to consume unbiased content, as most information is polarized.
These factors can lead to the formation of echo chambers within society, driving conversations away from consensus-building discussions.
To combat both idle reading and media bias, researchers have proposed several reading support tools.
Idle reading support tools \cite{Fok2022ScimIS,10.1145/3589955,Kang_2023} support highlighting, critical thinking, question answering, etc for increasing user attention span and understanding of the text.
Several research projects and commercial websites have tackled media bias by providing multiple perspectives \cite{10.1145/1518701.1518772,10.1145/3379350.3416154,10.1145/3383583.3398561} or guiding users to the same article written from the other side \cite{AllSides, GroundNews}.
However, these approaches require the user to take the additional effort to read another article, which users don’t always have the time or motivation for.
Moreover, there is a delayed effect, where the user could already be influenced by one article before reading the next one. \cite{doi:10.1073/pnas.2020043118} shows that timing is often the most critical part of fact checking. Hence, we argue that users benefit from seeing counter-arguments next to the arguments in real time as they are reading an article. 

To address these issues, we present ArguMentor,\footnote{We will release the ArguMentor system as open source software and the articles used if this paper is accepted.} a human-AI collaboration system that enriches the reading experience of opinion pieces by highlighting claims and generating counter-arguments to help users formulate more robust, nuanced views on important societal issues.
ArguMentor provides two types of support: passive and active.
The system provides \textit{passive} support by automatically highlighting main claims in the original text, generating counter-arguments for them, and providing a context based summary for the overall text.
ArguMentor provides \textit{active} support by allowing users to interact with the system by accessing the Q\&A bot (that answers user questions pertaining to the text), a DebateMe feature (an agent that users can argue any side of the piece with), and a highlighting trigger window (where users can highlight a word or passage for its definition or context).
We derived these features from a preliminary study of news readers, which illuminated key challenges and potential features and informed our design goals.

We evaluated ArguMentor via a within-subjects experiment (N=24). We found that, compared to the baseline, ArguMentor was able to help participants generate more claims and counter-arguments, boost their memory for those claims, and even gain better critical thinking skills. Follow-up interviews showed how this system aids those with lesser vocabulary or reading skills, where reading an article is more challenging. 
In summary, the contributions of this paper are as follows:
\begin{itemize}
    \item We propose an opinion article reading tool, ArguMentor, that augments the user experience with tailored AI-generated counter-arguments, a debate bot, and a question answering bot.  
    \item We demonstrate the effectiveness of ArguMentor via a within-subjects experiment with 24 participants.
    \item We provide insights and design considerations for future similar tools to enhance critical thinking.
\end{itemize}

\section{Related Work}
\subsection{Paper Reading Support Tools} 
HCI researchers have proposed a variety of software tools and technologies to support reading comprehension and critique of academic papers. 
To support reading comprehension, Kim et al.~\cite{10.1145/3242587.3242617} created an interactive document reader that links the text with its corresponding table cells automatically, which can reduce split attention and facilitate reading. 
August et al.~\cite{10.1145/3589955} created Paper Plain, a system that utilizes NLP techniques to enhance understanding of medical papers. 
Fok et al.~\cite{Fok2022ScimIS} created a skimming tool that highlights specific parts of the text in different colors to guide the reader’s attention, and enable them to read more efficiently. 
Kang et al.~\cite{Kang_2023} makes the process of finding other papers related to the current paper the user is reading easy by creating a citation graph and threading and summarizing their content using GPT-4. 
Other recent work \cite{10.1145/3490099.3511162,10.1145/3491102.3517470} adds on to this process by automatically highlights important citations, and provides commentary for them based on these papers, and creates a social network for various papers.

To support close reading and critique of written work, 
\cite{10.1145/2883851.2883965} propose a web-based collaborative platform that supports peer interactions and provides feedback for both students and teachers to engage in critical paper reading together.
Crebot~\cite{10.1016/j.ijhcs.2022.102898} asks users critical questions as they are reading the passage to further their understanding of it. 
Critrainer~\cite{10.1145/3586183.3606816} uses text summarization techniques and template based questions to help users raise critical thoughts. 

Our work builds upon these advancements by incorporating key features such as highlighting important points, summarizing content for better user understanding, and developing a question-answering framework. By adapting these established techniques to the domain of reading opinion articles, we aim to address the unique challenges associated with engaging critically with one-sided narratives, promoting a more balanced and comprehensive understanding among readers.

\subsection{News Reading Support Tools} 

\subsubsection{HCI and News:}
Extensive research has been conducted to enhance news consumption through novel technologies and interfaces. Laban et al.~\cite{laban2023designing} design and evaluate news reading interfaces that incorporate discord questions to reveal coverage diversity. 
Chen et al.~\cite{chen2023marvista} developed Marvista that employs Natural Language Processing (NLP) techniques like abstractive summarization to provide text-specific assistance when users are reading online articles. Their main evaluation study showed that Marvista helps users better comprehend the article. 
Nguyen et al.~\cite{10.1145/3242587.3242666} blends information retrieval and human knowledge to create a fact checking portal that aids human fact checkers.

\subsubsection{Combating Media Bias using HCI tools:} 
NewsCube~\cite{10.1145/1518701.1518772} provides readers with multiple perspectives on the same news, by showing them the article written by multiple sides. 
AllSides~\cite{AllSides} also does a similar thing by ranking media outlets, and showing a right, left and centrist perspective on the news. 
Perez et al.~\cite{10.1145/3379350.3416154} propose a browser extension that presents different perspectives by recommending articles relevant to the current topic.
Munson and Resnick~\cite{10.1145/1753326.1753543} evaluate the extent to which highlighting user agreeable terms within text or showing them first has an impact on their opinion. 
Hamborg et al.~\cite{10.1145/3383583.3398561} evaluate word choice bias in media by highlighting trigger words like “terrorists” etc and indicating their positive or negative sentiment. 

Our work differs from these efforts in two key ways. First, ArguMentor allows the user to simultaneously read the article and its counter-argument instead of referring them to another source. 
We expect that this quicker presentation has a beneficial effect on the user in that they are more likely to read the counters, and less likely to be influenced by the article in the first place. 
Second, ArguMentor refutes the argument from the article using an LLM-generated counter, instead of another (potentially biased) human-written article from an opposing viewpoint. This allows us to provide a succinct, relevant, contextualized counter-argument rather than directing the reader away from the current opinion piece in order to read additional opinion pieces. 

\subsection{Integration of LLMs to persuade users} 
The use of Large Language Models (LLMs) for tasks related to persuasiveness has been explored in several other contexts. 
Hyben et al.~\cite{hyben2023bigger} tested LLMs and fine-tuned models for claim detection to tackle things like misinformation and spread of bias. 
Khan et al.~\cite{khan2024debating} show that debating with persuasive LLMs leads to truthful answers and show that debate with an LLM is a good way to resolve conflict in cases where ground truth is unavailable. 
Breum et al.~\cite{breum2023persuasive} show that chatbots and LLM agents can generate powerful and persuasive arguments, and how they can play an important role in online discussions. 
Argyle et al.~\cite{doi:10.1073/pnas.2311627120} uses chatbots to show that online political conversations can be improved with an AI assistant’s suggestions. 
Karinshak et al.~\cite{10.1145/3579592} demonstrate that AI can be used to create effective public health messages, and that people are often persuaded by messages created by LLMs. 

These prior works lay the groundwork showing that LLMs have the potential to be persuasive and change users' opinions. We use LLMs in a similar way (by leveraging them as chatbots and prompting agents), but adapt and extend these ideas to the novel context of an interactive system supporting engagement with opinion pieces.

\section{Preliminary Study}

To support the creation of this system, we conduct an initial survey to gauge the challenges people face when reading opinion pieces and to better understand user attitudes and preferences towards potential design features.

\subsection{Potential List of Features} 
To brainstorm the potential features a reading experience aiding system could have, we reviewed the aforementioned recent literature in paper reading tools (e.g., \cite{Fok2022ScimIS, 10.1145/3491102.3517470}), along with creating a survey asking users for their feedback on a potential list of features.
The final list of potential features that is presented to participants of the survey is shown in Table~\ref{table1}.

\subsection{Survey Method} 
We recruited 40 participants to complete our survey. The age range was 20--45. Fifty-five percent were men and 45\% were women. For 30\%, English was their second or higher language. There was a diverse mix of educational backgrounds, from those who did not complete high school, to college graduates, to those with advanced degrees. Around 70\% of them were frequent news readers. These participants were selected through snowball sampling participants at our university and an online gaming group (these groups tend to attract a diverse range of individuals from various backgrounds). The questions focused on their demographics and news reading habits, information about their interaction with opinion pieces, and design ideas for potential technological support. 
Details of the survey protocol and participant backgrounds can be found in the Supplementary Materials.

\subsection{Findings}
\subsubsection{Challenges of Reading Opinion Pieces} 
\begin{itemize}
    \item \textbf{C1: Difficulty Focusing and Completing the Articles:} A majority of the participants indicated that they struggle with long opinion pieces because it is difficult to stay engaged in a long piece. For example, P3 wrote, “Opinion pieces are often really long, and I find myself reading a passage, `dozing off, then having to re-read it.”
    \item \textbf{C2: Confirmation Bias:} Consistent with prior work, a few participants note that they struggle with confirmation bias, where they much prefer reading articles that support their point of view, and avoid reading articles from the other side altogether. For example, P1 noted, “Sometimes its difficult to avoid the tunnel vision that you experience when arguing for or against a topic in a specific way." 
    \item \textbf{C3: Difficulty Imagining Counter-arguments:} A few of our participants also noted that they have a hard time thinking about what the other side might say, when they are invested in reading an opinion piece. They indicate that they often have strong opinions about the current piece, one way or another, and can’t easily dismiss those thoughts. When surveyed on whether they go back to researching the other side once they’re done reading, a majority of these participants indicated that they don’t, and often the opinion from the article sticks with them. For example, P1 responded, "It becomes difficult to think of creative ways to defend/oppose the argument from the templated way in which you are used to doing so.” 
\end{itemize}

\subsubsection{Usefulness of potential features} 
Participants were asked to rate potential features on a scale of 1--5 (5 = most useful). 
Table~\ref{table1} shows the results for this part of the survey. 

\begin{table}[h]
    \centering
    \begin{tabular}{|c|c|c|}
        \hline
        Potential Features & Mean & StdDev \\
        \hline
            \textbf{Highlight main arguments} & \textbf{4.33} & \textbf{0.79}\\
            \textbf{Counter-arguments} & \textbf{4.24} & \textbf{0.88}\\
            \textbf{Current events context} & \textbf{3.95} & \textbf{0.86} \\
            Summary & 3.89 & 1.08 \\
            Left vs Right comparison & 3.71 & 1.10 \\
        \hline
            \textbf{DebateMe} & \textbf{4.19} & \textbf{0.98} \\
            \textbf{User style adoption} & \textbf{3.82} & \textbf{0.97} \\
            \textbf{Highlight for context} & \textbf{3.80} & \textbf{0.93} \\
            \textbf{Q\&A} & \textbf{3.71} & \textbf{0.85} \\
        \hline
    \end{tabular}
    \caption{Feature Survey Results}
    \label{table1}
\end{table}

\subsubsection{Open-ended questions}
Our participants also responded to open-ended questions about other potential features they would like to see. 
P18 suggested that they would like to see the main claims from the passage highlighted right next to the counterclaim, which is also highlighted in the same color. 
A few participants suggested that the “DebateMe” feature should take feedback about persuasion from the user, and modify its arguments accordingly.
Similarly, some participants suggested that the counterarguments should also take user feedback for whether they are persuasive. 
The final system was designed based on feedback from this initial survey, and will be described in the following section.

\section{System Description}


Informed by the preliminary study and our literature review, we created ArguMentor, a human-AI collaboration system that enriches the reading experience of opinion pieces. In this section, we present a user scenario, followed by a description of the back-end architecture.

\subsection{User Scenario}
In this user scenario, we describe how John, a frequent news reader, would use this system. John begins by providing an opinion piece about monetary policy to ArguMentor that he wants to read. Next, he will see the op-ed text with claims highlighted on the left, along with counter-arguments for those claims on the right. He has limited time to read right now, so he clicks on the "get extra context" button, which generates a summary and provides broader context from online information. Then he skims through the counter-argument summaries. 
  
Later, John uploads a second opinion piece on the Trump indictment when he has more time to read. He's less familiar with this topic and isn't sure where he stands on the issue. While reading, John encounters a legal term, "statute of limitations," so he uses the Highlight feature to select that text and request a contextualized explanation. Then he uses the Q\&A feature to type in a question about the demographics of the cities mentioned in the article. Finally, he feels comfortable he understands the author's key points. He activates the DebateMe feature to express a key belief he shares with the author; the bot plays Devil's Advocate and provides an opposing viewpoint in real time. 

Having completed the interaction, John now has a better understanding of not only the opinion piece, but also of where he stands on this topic. 
He can also argue for it better in the future. 
In summary, John can passively or actively interact with the system to ensure that he is not influenced by just this article, and knows the full context of the piece. 

\subsection{System Architecture}
\begin{figure*}[h!]
    \centering
    \includegraphics[width=0.9\textwidth]{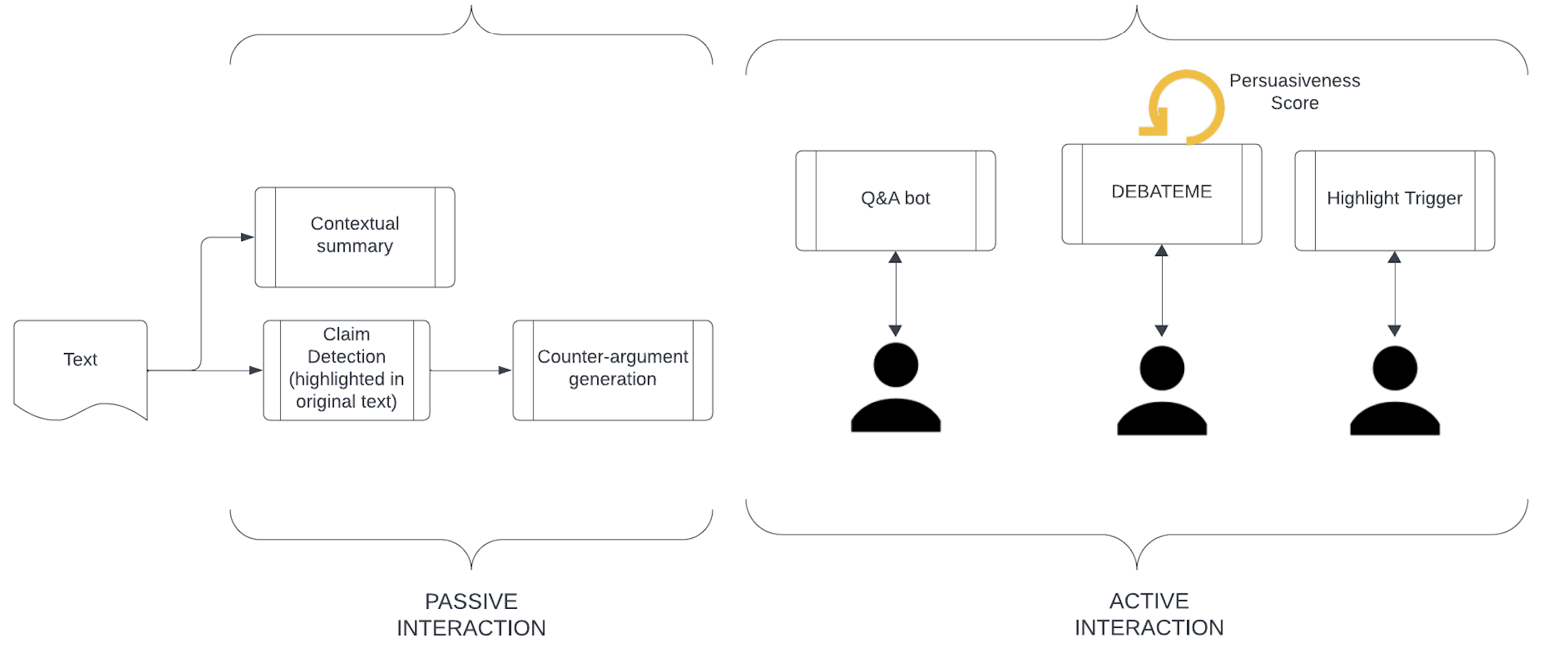}
    \caption{Architecture Diagram}
    \label{figure1}
\end{figure*}

ArguMentor is deployed on Vercel \cite{Vercel}, and Mouseflow \cite{Mouseflow} is used on the website to track user activity and record sessions. 
The system is divided into 2 stages: passive interaction and active interaction (see Figure~\ref{figure1}).

\subsubsection{Passive Support:} 
There are three forms of passive reading support: context of the article, highlighted main claims in the article, and counter arguments for those claims (see Figure~\ref{fig:screenshot}). 
\begin{enumerate}
\item \textbf{Summary/Context}:  Upon clicking a button that says “Get Additional Context”, users can get a neutral context about the article from the internet. This is done using SerpAPI \cite{SerpAPI}.
GPT-3’s knowledge is limited to 2021, as of this paper, and hence getting context about recent articles is not possible just using GPT-3 API. 
Using SerpAPI and its “zero-shot-react-description” agent can be used to get direct google search results. 
The prompt passes the title of the article and asks to summarize the context of this issue. 
This achieves two purposes: gives the readers a quick summary, and ensures that the summary is not biased, but rather based on the context of the passage. 
Users in our preliminary survey indicated that they preferred this summary over a summary of the article, hence it has been implemented. 

\item \textbf{Claim detection}: Claims are automatically detected within the text and highlighted. 
A fine-tuned model is used for claim detection.
The GPT-3.5-turbo model is fine-tuned using the IBM-30K claims dataset \cite{aharoni-etal-2014-benchmark}. 
First, we structured the dataset according to the fine-tuning guidelines of OpenAI, where the system instruction was to return claims as it is, and user and assistant instructions were examples from the dataset.
Only 11 instances are able to produce a good result for fine-tuning a large model like GPT3.5 \cite{OpenAI}.
Once the model is able to return claims from the passage, the claims are matched to the original text using REGEX.  An instance of GPT-3.5 cannot be used by itself because we need to return an exact match so it can be highlighted- this is something a fine-tuned model achieves significantly better.
The match is then highlighted in yellow. 
The main challenge in this part was to create the fine-tuning dataset such that it would account for a diverse range of news articles of various lengths.  

\item \textbf{Counter-argument generation}: To generate effective counter-arguments, we fine-tune a GPT 3.5 based model using a rebuttal dataset \cite{orbach2019datasetgeneralpurposerebuttal}. The model was trained on 32,176 tokens across 8 epochs with a batch size of 1 and a learning rate multiplier of 2. The final training loss is shown in Figure~\ref{fig:training-loss}. LLMs by themselves are shown to generate effective rebuttals, but fine-tuning a model on a specific dataset ensures that the rebuttals are even more effective. \cite{carrascofarre2024largelanguagemodelspersuasive} show that these rebuttals, generated by LLMs are targeted, specific and persuasive, which would achieve a better result than just in-context learning. 
These counter-arguments are then displayed on the right hand side of the page. The users can click on "Expand" to see the whole counter-argument if they are interested in it. 
Although most times the contexts are efficient and refute the claim, sometimes they can be vague, especially if the claim is about a recent event that the LLM is not trained on. 

\begin{figure*}[h!]
    \centering
    \includegraphics[width=0.6\textwidth]{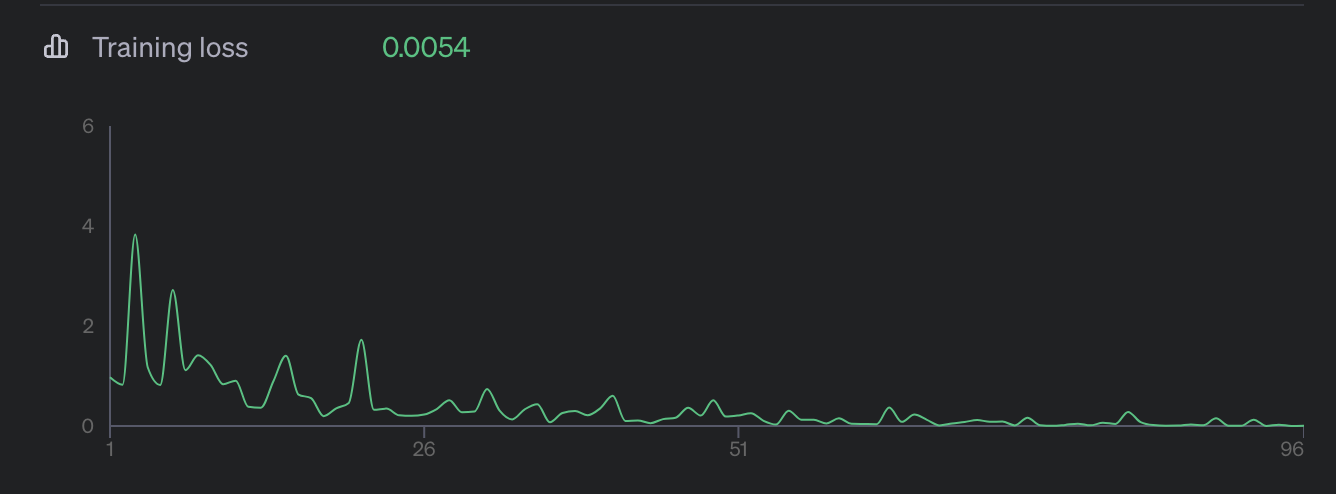}
    \caption{Training Loss: OpenAI GPT 3.5 training loss by training over the IBM Rebuttals Dataset}
    \label{fig:training-loss}
\end{figure*}

\end{enumerate}

\subsubsection{Active Support:} 
\begin{figure*}[h]
    \centering
    \includegraphics[width=0.8\textwidth]
    {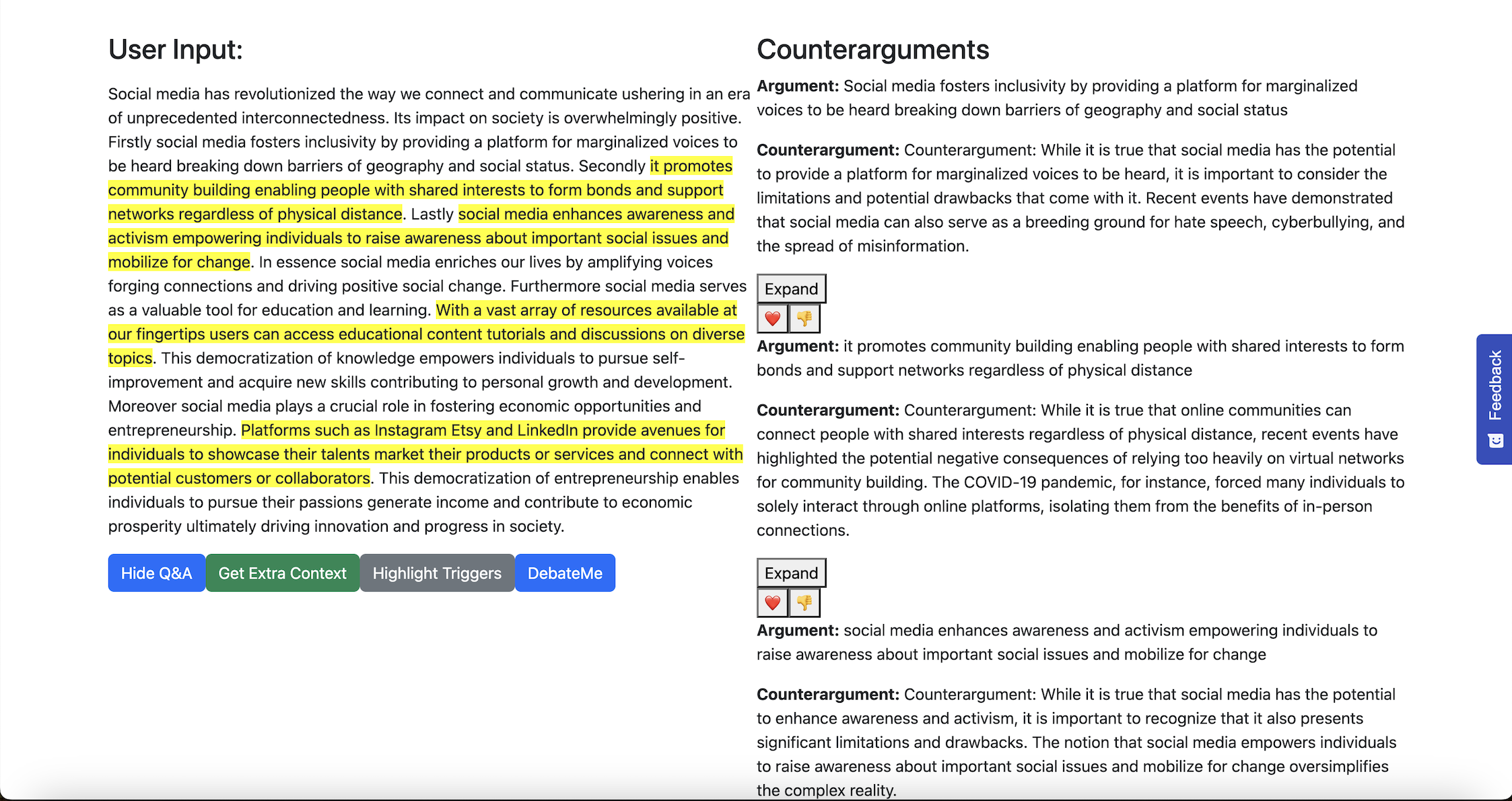}
        \caption{ArguMentor System Screenshot}
        \label{fig:screenshot}
\end{figure*}

Active support (interaction) comes primarily in three forms: Q\&A, DebateMe, and highlighting (see Figure~\ref{fig:screenshot}). 
\begin{enumerate}
\item \textbf{Q\&A}: There were two options for creating a Q\&A agent: a prompting method like counter-argument generation, or a RAG architecture.
A RAG architecture is less likely to hallucinate, provides more relevant responses, and reduces biased responses \cite{gao2024retrievalaugmented}.
Hence, the latter was chosen for this architecture. 
OpenAI embeddings were used to convert the user query and the document that the user has uploaded into a vector space, and store it in the Chroma vector store.
A button was created that initiated a chatbot where users could type their query and get efficient responses from the bot.
History of the conversation is also shown on screen.

\item \textbf{DebateMe}: This is implemented using LangChain’s conversation chain which enables chains of conversation. 
Similar to \cite{DebateDevil}, the user can enter their argument, and the bot will debate the other side.
This is implemented using prompting --- the LLM is asked to debate the other side to the user’s argument, and be as brief and persuasive as possible.
The user is then given an option to click on a "thumbs up" or "thumbs down" button. If the user clicks on "thumbs down", the LLM will find another way to persuade the user. This is done to ensure that the user is satisfied with the response, and can report any problematic content to the developers. This is also done using in-context prompting. 

\item \textbf{Highlighting window}: A window is enabled where the user can highlight specific parts of the text to get more information about it, including definitions, counter-arguments, additional information, etc. The text selected by the user is sent along with the following prompt: "If it's one word, provide its definition. If it's more than that, use your knowledge to give additional context on the text".

\end{enumerate}

\section{Evaluation}
To investigate the benefits of ArguMentor for users reading opinion pieces, we conducted a within-subjects experiment with 24 participants. In this experimental design, each participant read one op-ed using ArguMentor (experimental condition) and one without the system (control condition), i.e., typical reading without any intervention. 

\subsection{Baseline}
We select reading the article without any intervention (no system, or any other support tool) as our baseline. We justify the choice of our baseline here: 
\begin{itemize}
    \item First, most readers in our preliminary survey indicated that this is how they consumed news on a regular basis. We know this to be representative of the general population as well. In that case, it makes the most sense to select this as the baseline. 
    \item Secondly, even though websites like GroundNews and AllSides exist, they work very differently to our system. For example, ground news categorizes a news piece as left-, right-, or center-leaning. It then identifies how that issue has been covered by all three sides and various outlets. There are no counter-arguments, other than if the right covered a topic completely different from the left. On the other hand, our system provides direct rebuttals to a given claim. It does not make the users go through another article written from a different viewpoint. 

\end{itemize}
\subsection{Articles}
To ensure a diverse selection of op-ed articles, we chose six articles from the news spanning three subfields: politics, social issues, and economics. For politics, we focused on the U.S. election as a key topic; for economics, we selected stock buybacks; and for social issues, we focused on feminism. This range represents a broad spectrum of socio-economic topics, allowing us to evaluate the effectiveness of our system across different subject areas. Two articles are chosen per key topic, and each article can be read with our system or with the baseline (without the system), creating 12 unique conditions. Participants weren't required to know about the topics in order to participate in the study. 

\subsection{Task and Procedure} 
\begin{figure*}[h!]
    \centering
    \includegraphics[width=1\textwidth]{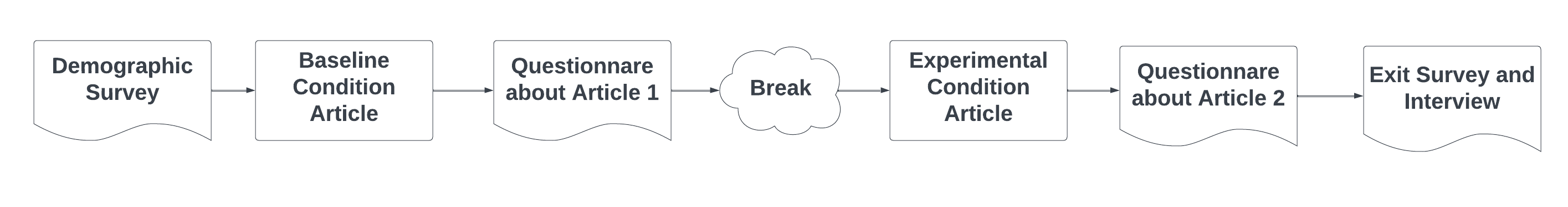}
    \caption{Experiment Design and Procedure: The process always starts with the baseline condition (reading the article by itself), and ends with the experimental condition (reading the article with the system). Each person only reads two articles and submits questionnaires right after reading.}
    \label{figure2-exp-design}
\end{figure*}

The experiment is conducted in several parts. The entire procedure is shown in Figure~\ref{figure2-exp-design}.

First, demographic information is collected for all participants. Then, participants are given the baseline condition. The baseline condition is always presented first because it avoids participants finding out what the ArguMentor system (experimental condition) looks like and attempting to replicate its functionality later for the baseline. 

Next, participants are presented with a survey. The survey asks several questions. First, it asks attention checking questions to see if the participants understood the basic facts in the article. These are multiple choice questions such as what was the topic of article, who did the article support, what is the main point of the article, etc. 

The survey then asks participants to write any claims, counter-arguments, questions, biases, and solution to problems they can think of, based on the article. They are then asked to answer ACT-style questions related to deduction, assumptions the author makes, whether the author should remove or keep certain sentences to prove their point, etc. These questions are generated by prompting a Llama-3 model to generate ACT style questions based on a paragraph. The authors have then gone through each question in detail to ensure they are answerable and valid. The claims and counter-arguments are timed, whereas the critical thinking portion is left untimed. This is because critical thinking can take some time, and we give participants the flexibility to take breaks, etc to think of the given questions/prompts. We time claim and counter-argument generation to see the immediate effectiveness of our system. The choice for selecting these metrics is explained in the next section.   

This procedure is then repeated with the experimental condition, with participants getting access to ArguMentor. A final, exit survey consisting of subjective experience and system feedback questions is given in the end. Participants can have as much time with each article and the system as they wish. They are instructed to not return to the article or the system while filling the survey, and answer subjective questions as honestly as possible. 

The document selection is counterbalanced to reduce order effects; i.e., Participant 1 gets Document A in the control condition and B in the experimental condition, Participant 2 gets Documents B in the control condition and A in the experimental condition, and so on. However, all participants experienced the experimental condition after the control condition. 24 people read 2 articles each, resulting in 48 data points for our 12 conditions (6 articles, each with and without the system).

\subsection{Evaluation Metrics} 
We measured participants' outcomes, subjective experience, and processes to see how interaction with ArguMentor compares to the baseline. 

\subsubsection{RQ1: Outcomes} 
We measured participants' outcomes in two areas: (1) finding claims and (2) thinking of counter-arguments for those claims.
As our preliminary study indicated, users have a hard time doing both of these tasks, so we want to see if the system improves these skills. These two skills are also fundamental for everyday tasks; i.e., being able to spot the main claim in a news article, social media post, or a live debate and then thinking of a potential counter to it to see what the other side would say. 
We count the number of claims / counterarguments participants submit in each post survey. Our hypothesis is that use of ArguMentor will increase the number of both.

We also measure how many biases present in the article users can write about, how many questions they pose, how many solutions they pose to the problem in the article, and how many ACT style questions they answer correctly (out of 10). This is the critical thinking portion of the experiment. These are the facets of critical thinking tests such as \cite{Critical1} as described in \cite{aurl}. The Watson Glaser Critical Thinking Test measures "RED" -- Recognize Assumptions, Evaluate Arguments, Draw Conclusions through various steps. We want to see if our system erodes or promotes critical thinking. The system only gives them counter-arguments, and the ability to ask questions related to the article, how creatively they use it for these tasks, and whether the ultimate outcome is better are metrics to be evaluated.

\subsubsection{RQ2: Process} 
We observe the process by seeing how long the participants spends on which feature, and how that contributes to the overall number of arguments.
We also measure the number of arguments the participants come up with and divide that by the time they spend on the baseline or system to normalize it, and see the effect the system has on creating any additional burden for the user.

If the participants can come up with an equal number of arguments for both, but the system takes them twice as long to use, it might suggest that the system is burdensome to use. The benefits are clearer if the participant can come up with more counter arguments after using the system for more or less an equal time than just reading the article. However, it is important to consider that in contexts where efficiency is not a priority, such as leisure reading, taking more time might be acceptable if the system offers a better subjective experience. We also want to measure how the user spends time on the system: which features they use, for how long, etc.

\subsubsection{RQ3: Subjective Experience} 
We measure the participant’s subjective experience of the system by asking them how they feel about the system. We adopt the technology acceptance model as well as a survey to collect feedback on the system. A short interview is also conducted to allow them to express their genuine thoughts. 

\subsection{Quality Analysis}
Participants might just generate more arguments with the system, but that doesn't convey how strong these arguments are. To gauge this, we compare the quality of counter-arguments the participants generate with the baseline. This is done in two ways: computational and human. Increasingly, research has shown the prominence of GPT-as-judge \cite{zheng2023judgingllmasajudgemtbenchchatbot}. We deploy this method to ask a model the binary question, "which [solution/argument] is stronger: x (baseline) or y (system)?". This is compared across topics and participants. To avoid any potential bias such as a model preferring an argument that sounds like its own, we use a Llama-3 7b model for this task. 

Secondly, we also recruited three expert debaters, who have judged and "broken in" (a debate term for qualifying for out rounds such as semi-finals or finals) over 3 international parliamentary debating competitions such as UADC, Australs, etc. We ask them the same binary question about a randomly selected counter-argument from each survey we collect (each expert answered 48 questions). These counter-arguments are presented in random order (from the baseline or the experimental condition). We check their agreement on this task, and only consider the argument as superior if all 3 experts voted for it. In case of a non-unanimous answer, we ignore that data point.    

\subsection{Participants} 
We initially recruited 5 participants from a gaming convention group. Each was then instructed to recommend another person for the study (snowball sampling) until a total of 24 participants were recruited. Eleven participants were men, 11 were women, and 2 identified as other genders. 33\% spoke English as their first language, while a majority spoke at least one other language. Most of them reported reading the news at least semi-regularly (twice a week). The participants ranged from 19--43 years old, representing a vast range of individuals. They also represented a variety of income brackets, with over half being middle or lower-middle class. Ten had only finished high school or had not finished high school, 8 had a bachelor's degree, and 6 held advanced degrees. 

\section{Results}

We perform the Mann--Whitney U test~\cite{10.1214/aoms/1177730491} to compare the difference with and without the system for all research questions.  The Mann--Whitney U is a non-parametric test commonly used to compare differences between independent conditions especially when the data normality is violated, as confirmed in our cases. In all U tests, we set the significance level at the standard threshold of $a$=0.05.

All 24 participants scored 100\% on all the attention-checking questions.

\subsection{RQ1: Outcome Results} 
\subsubsection{Number of claims:} Figure~\ref{fig:NumClaims} shows the number of claims increase by at least two times when the users engage with the system. These differences were significant (p=0.01, 0.001, 0.005). The overall word count also increased by over 200\% for the claims.  
Along with the topic, we also averaged each participant's improvement using the system over the baseline. A percent increase of 170\% is observed overall, which shows that our system helped each individual as well. 

\subsubsection{Number of counter-arguments:} Similar to claims, the number of counter-arguments users were able to generate increased significantly, ranging from a 0.6x increase (economics) to a 2.5x increase (social) (p=0.01, 0.004, 0.02) for economics, politics and social, respectively). Figure~\ref{fig:NumClaims}  shows the number of counter-arguments with and without the system. The overall word count for the counter-arguments again increased by over 150\% over the baseline. Similar to claims, counter-arguments saw an individual percent increase of 150\% across all topics.

\begin{figure}[t]
    \centering
    \includegraphics[width=0.5\textwidth]{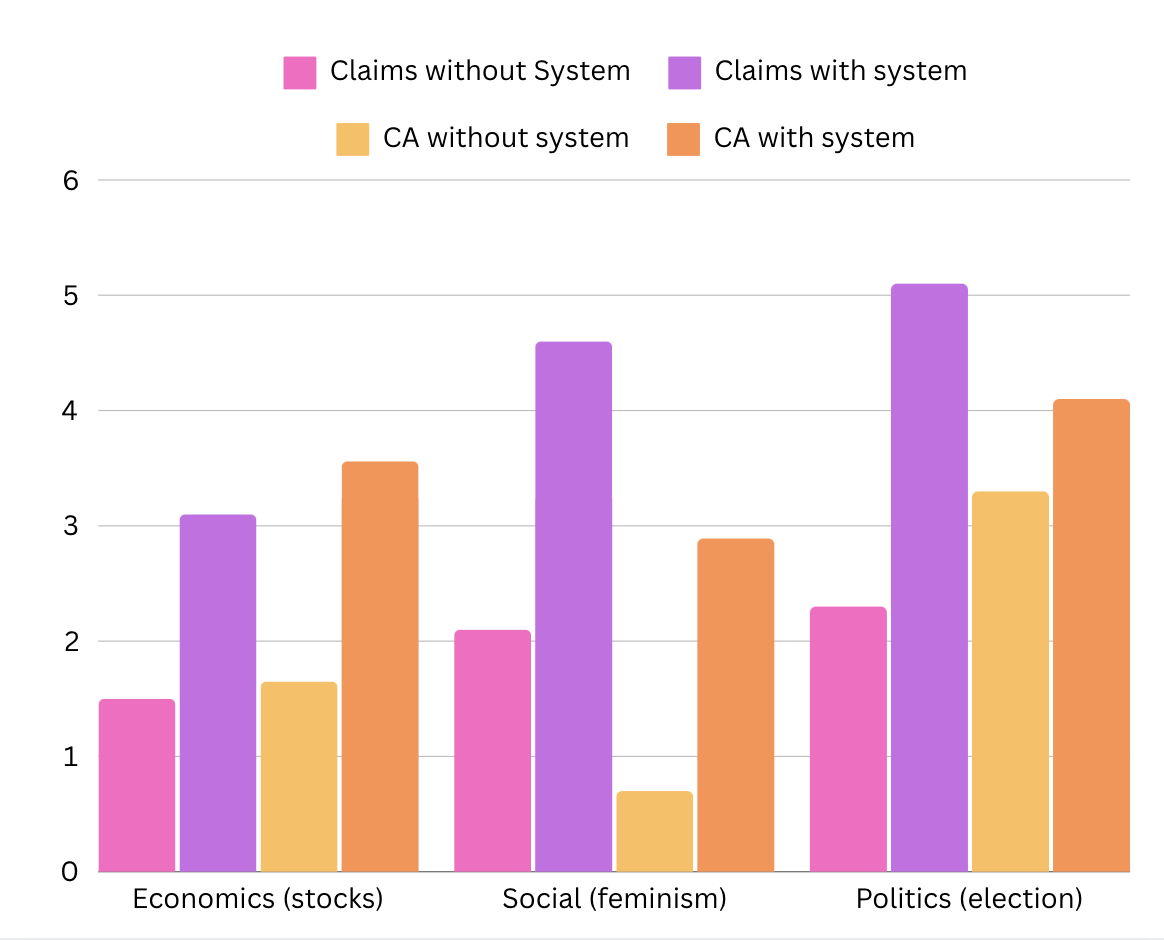}
    \caption{Performance: Number of Claims and Counter-Arguments with and without ArguMentor.}
    \label{fig:NumClaims}
\end{figure}

\subsection{Critical Thinking Skills}
Figure~\ref{fig:NumClaims} shows how various critical thinking skills changed with the baseline. Biases, questions and solutions were judged, and all saw an increase with the system, even though the system doesn't directly do any of this for the user. For the ACT style questions, participants answered, on average, 60\% of the questions correctly without the system, and 78\% of the questions correctly with the system, showing another indirect benefit of the system. 

\begin{figure}[t]
    \centering
    \includegraphics[width=0.5\textwidth]{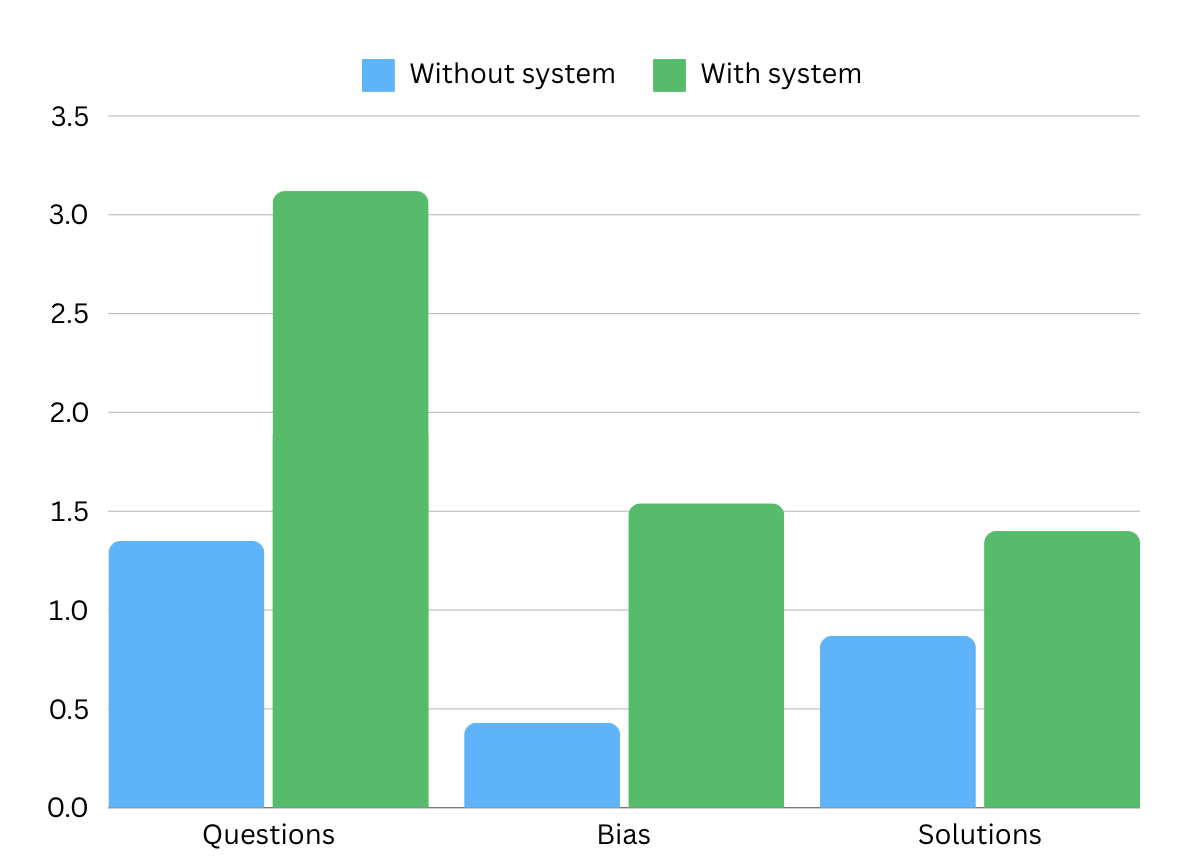}
    \caption{Performance: Average Critical Thinking Skills over all three domains}
    \label{fig:NumClaims}
\end{figure}

\subsection{Quality of Arguments}
The computational quality assessment (Llama-3 7B) preferred the counter-arguments generated by participants after they used ArguMentor 83\% of the time. For the human experts, 35/48 instances showed unanimous agreement and hence were kept. Out of this, experts preferred counter-arguments generated by participants after they used the system 71\% of the time. This shows that both human experts and AI-as-judge agreed that arguments were higher quality after the participants experienced the system. 

\subsection{RQ2: Process Results}
\subsubsection{Claims and Counter-arguments per minute:} 
\begin{figure}[t]
    \centering
    \includegraphics[width=0.5\textwidth]{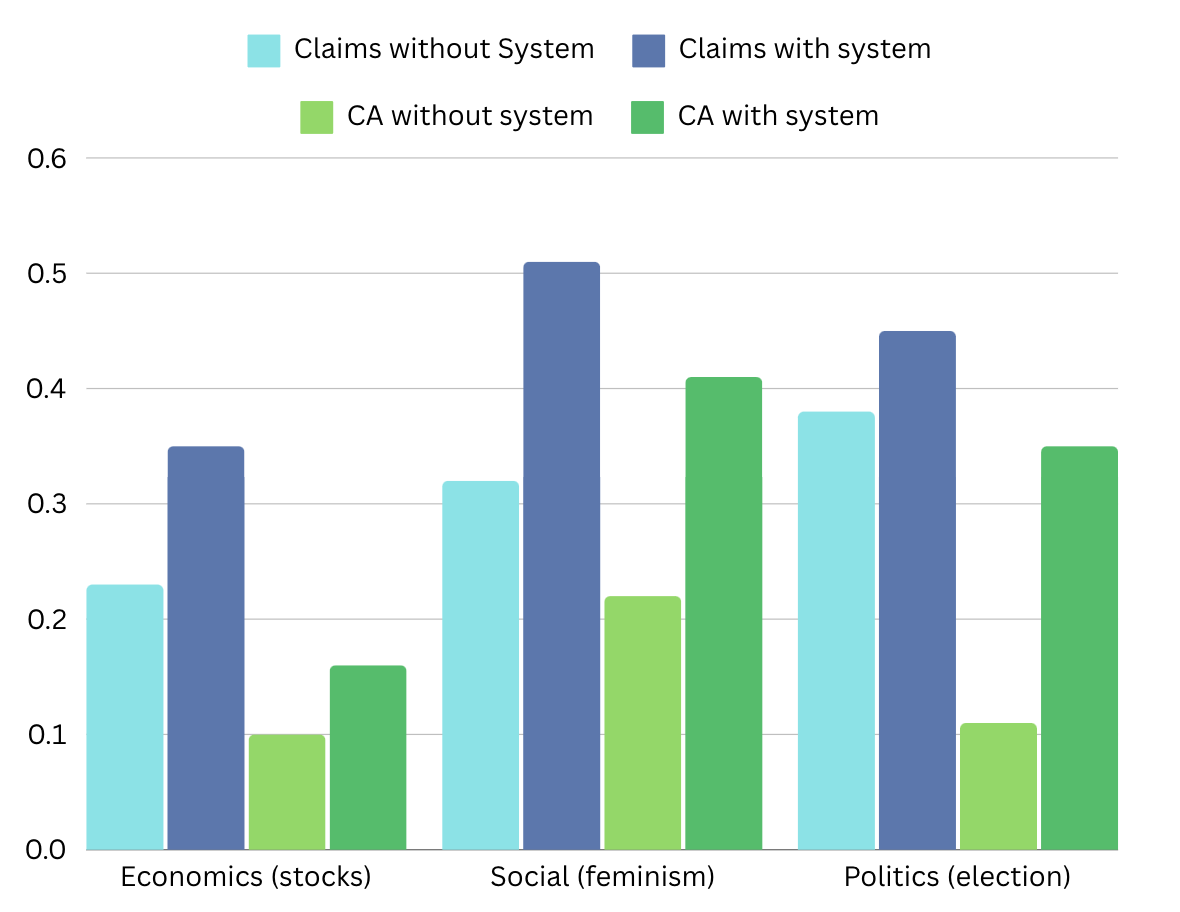}
    \caption{Process: Claims and Counter-arguments per minute with and without Argumentor}
    \label{fig:ClaimsPerMinute}
\end{figure}

Figure~\ref{fig:ClaimsPerMinute} shows how claims per minute increase for all types of articles when participants use ArguMentor. Counter-arguments per minute are similar for the baseline and the system. 

\subsubsection{Time spent on system by activity:} 
\begin{figure}[h]
    \centering
    \includegraphics[width=0.35\textwidth]{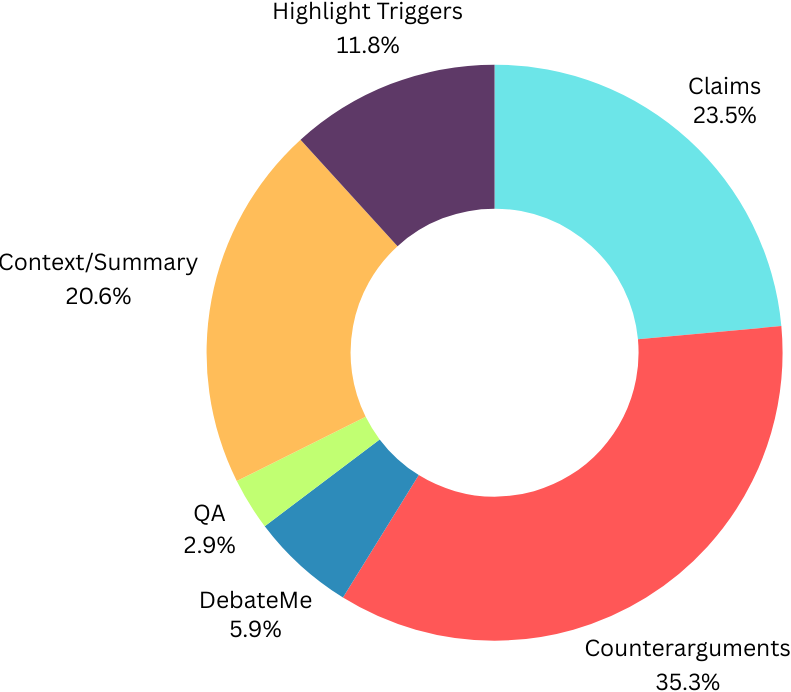}
    \caption{Time Spent using ArguMentor features.}
    \label{fig:TimeSpent}
\end{figure}

Figure~\ref{fig:TimeSpent} shows how long participants spend on which feature in the system. This data is acquired using Mouseflow logs. Over half the time was spent on claims and counter-arguments, while the Q\&A and DebateMe features were least used with less than 10\% of overall time spent. These results align with the preliminary study's survey results in Table~\ref{table1}, where participants used main claims and counter-arguments the most.  

\subsection{RQ3: Subjective Experience Results}

\subsubsection{System experience:} Table~\ref{tab:Subjective_Opinions_Feedback} shows how the users perceived the system and its various features. The system scored high for helpfulness and ease of use. When asked in detail, users consistently reported DebateMe or Q\&A features being the least helpful, consistent with the time spent analysis. Participants cited lack of interest in the articles, and the process being too time consuming as the primary reasons for not using these features. However, participants who found an article particularly interesting enjoyed the DebateMe feature.

\begin{table}[ht]
    \centering
    \begin{tabular}{|c|c|c|}
        \hline
        Factor &  Mean & SD \\
        \hline
         Ease of Use & 4.71 & 0.46 \\
         Motivation to Use & 4.50 & 0.51 \\
         Helpfulness & 5.00 & 0.00 \\
         Frustration (inv.) & 4.13 & 0.68 \\
         Mental Demand (inv.) & 3.3 & 1.07 \\
         Recommend to friend? & 3.58 & 0.97 \\
         Daily Use & 2.88 & 0.90 \\
         Persuasiveness & 4.13 & 0.90 \\
         \hline
    \end{tabular}
    \caption{Subjective Results: System Feedback}
    \label{tab:Subjective_Opinions_Feedback}
\end{table}

\section{Discussion}

\subsection{Revisiting the Research Questions}
In this paper, we propose ArguMentor, a system designed to help users gain a more nuanced, balanced understanding of news opinion pieces. We note that users often encounter one perspective when reading news articles, and could benefit from having access to counter-arguments. The larger goal of our system is to give users quick, easy access to multiple viewpoints while reading an opinion piece, helping them form a more informed, robust, and nuanced perspective. 

Addressing RQ1, we can see from the outcomes results that participants identified a significantly higher number of claims and counter-arguments with the help of the system. Without the system, most participants could only identify one main claim within the article. However, the system systematically highlights all claims throughout the article and provides them with specific counters, which can help users holistically understand the article. Participants were instructed not to return to the system or article when filling the survey, but were still able to write up to 3 times as many points after using ArguMentor, showing that the system of highlighting also helps with memory. This is consistent with other HCI systems which use highlighting to boost memory (e.g., \cite{Fok2022ScimIS}). This finding was noticed by participants as well; e.g., P4 mentioned how ``highlighting was the best source of the system. I didn't have to even read the entire essay, I got the gist of it quickly.'' 

We can also see that critical thinking skills rose across the bar for a variety of topics. Participants reported using creative methods like asking more questions, finding out all definitions, etc to identify biases, and pose solutions. They also reported using the counter-arguments to come up with a more nuanced solution to any problems in the article. In the exit interview, multiple participants talked about how reading the counter-arguments made them think of more innovative solutions to the presented problem. P11 talked about his process: "To detect biases, I mainly used the counter-arguments, then asked questions to the system about how the counter fixes issues in the claims". P20 talked about how reading counters made him think of his experiences and how both the claim and counter could be improved. An interesting thing we noticed is that most often users didn't repeat what the AI gave them as "rebuttals" when writing counter-arguments in the survey, rather adding their own details and thought process to it. This shows that a system like ArguMentor gets the user thinking about a potential counter. Most of our users echoed this, while P16 said "sometimes I thought the counter-argument was a bit weak, maybe it lacked recent nuance about the elections or something, but it got me thinking about a potential counters that apply to that too" 

In case of RQ2, we can see that the additional time participants spent on the system often led to a higher number of claims, and a similar number of counter-arguments. With counter-arguments the time it takes users to read the counter-argument presented by the system potentially adds to the delay, making it roughly equal to the time spent thinking using a baseline. However, users are able to generate longer counters with the help of the system (as indicated by a higher average word count), indicating higher confidence. Moreover, the quality of the counter-argument is higher, as indicated by both human experts and AI-as-judge. Beyond opinion pieces, news articles, especially about economics, science, politics, etc., can be complex, jargon-heavy, and unintentionally biased, indicating places where our system can provide additional value.

For RQ3, we found that ArguMentor was easy to use, helpful, and participants were motivated to use it. However, the recommendation for daily use is lesser because of the time it takes to not only read the article, but also read all the additional information the system provides. 

When the topic or article was particularly complicated, participants used ArguMentor's highlight feature (and benefited from it) more. The economics article on stock buybacks which was naturally harder and jargon-heavier than others, showed an increased usage of highlight features, and participants later reported finding those features useful as well. Similarly, even though DebateMe was used less overall, participants who were particularly passionate about a news topic used it (and found it useful) more for that article. Participants also reported finding the thumbs up and down features useful, as they were able to get several different arguments quite easily. 

\subsection{Impacts on Critical Thinking and Reading Skills}
An interesting question this raises, and that future systems should also take into account is whether such a system promotes or erodes critical thinking. On one hand, the results show that the facets of critical thinking (asking questions, identifying biases, writing counter-arguments) are all better with the system. On the other hand, there is not enough evidence to prove whether these are better because there is some inherent change in the ArguMentor user. This forces us to think about the role such a system plays in today's society. It fills in a few gaps for the users to fill in the rest. Argumentation and reasoning are skills that we develop over our lifetime, and such a system can help with critical reading on a regular basis to then encourage critical thinking even without the system. There is a risk, similar to the risk ChatGPT poses to our education system, where users get accustomed to system generated answers and stop thinking beyond that. To mitigate this risk, this system only gives users tools to think critically. We help them with a few potential rebuttals to claims they find in the passage, so that they can decide what they believe, what biases are present on both sides, and what solutions can come up with a compromise from both sides. We don't directly provide solutions or biases. Our participants echoed this experience in their exit interview. 

Moreover, for those less educated, this tool could provide value in developing critical reading skills. Our participants that had a high school education or less talked about how this system was different from their usual reading experience and how it felt like it aided them: P1 said "[ArguMentor] didn't shame me, didn't make fun of me for not knowing what, [for example] a stock buyback was --- that article was a tough read, and I got through it with all the definitions, highlights, I would love something like this everyday." Even if users completely ignored any counter-arguments or other features, they can simply read definitions of words in the passages and understand it better, something they probably wouldn't do otherwise. This tool, when used correctly, can enhance literacy on these topics, and reduce misinformation and bias. 



\subsection{Limitations and Future Work}
The goal of the system is to reduce the bias that articles bring in. 
However, since we are using LLMs to generate counterarguments and for chatbots, it is possible that they are either not as persuasive, not as relevant, or can carry more bias or misinformation compared to those written by expert humans. 
We have mitigated this by adding certain constraints to the prompt, and using models that have guardrails (like GPT-3), and using a RAG system rather than simply prompting wherever applicable. However, there is still a possibility that the model response has bias, and it should be monitored continuously. 

Even if poor arguments are generated, which aren't persuasive to users, at the least it offers users a chance to think of counter-arguments. It encourages the user to confront the possibility that the counter-argument provided is a poor one and that something else could potentially be better. 

\section{Conclusion}
In this work, we introduce ArguMentor, a novel human-AI collaboration system designed to enhance critical thinking and combat echo chambers in news consumption. ArguMentor highlights claims, identifies counter-arguments, and generates context-based summaries for the user. With active features like a question-answering bot and a debating agent, ArguMentor encourages users to engage deeply with the text. Our evaluation, conducted on a diverse group of participants, shows that users not only generate more claims and higher-quality counter-arguments but also raise more questions, identify biases, and propose solutions more frequently than when reading articles without the system. ArguMentor highlights a creative way in which AI can serve as a tool that boosts rather than erodes critical thinking, learning, and engagement with argumentation.

\bibliography{sample-base}

\end{document}